\documentstyle[12pt,blois,psfig]{article}
\def\ref{\par\noindent\hangindent=1truecm}
\font\piedi=cmr8

\catcode`\@=11
\def\gsim{\ifmmode{\mathrel{\mathpalette\@versim>}}
    \else{$\mathrel{\mathpalette\@versim>}$}\fi}
\def\lsim{\ifmmode{\mathrel{\mathpalette\@versim<}}
    \else{$\mathrel{\mathpalette\@versim<}$}\fi}
\def\@versim#1#2{\lower 2.9truept \vbox{\baselineskip 0pt \lineskip 
    0.5truept \ialign{$\m@th#1\hfil##\hfil$\crcr#2\crcr\sim\crcr}}}
\catcode`\@=12

\def\lsun{\hbox{$L_\odot$}}
\def\lb{\hbox{$L_{\rm B}$}}
\def\msun{\hbox{$M_\odot$}}
\def\IMLR{Fe$M/L$}
\def\micm{\hbox{$M_{\rm ICM}$}}
\def\mfecm{\hbox{$M_{\rm Fe}^{\rm ICM}$}}

\def\mfes{\hbox{$M_{\rm Fe}^*$}}

\def\zfes{\hbox{$Z^{\rm Fe}_*$}}
\def\zfecm{\hbox{$Z^{\rm Fe}_{\rm ICM}$}}
\def\ho{\hbox{$H_\circ$}}
\def\h50{\hbox{$\ho /50$}}
\begin{document}
\heading{HINTS ON GALAXY FORMATION FROM THE\\
         METAL CONTENT OF GALAXY CLUSTERS\\
          AND OTHER FOSSIL EVIDENCE}
\vspace{1cm}
\begin{center}
Alvio Renzini\\
{\it ESO, Garching b. M\"unchen, Germany}
\end{center}
\vspace{1cm}
\begin{bloisabstract}
Clusters of galaxies allow a direct estimate of the metallicity and
metal production yield on the largest scale so far. It is argued that
cluster metallicity ($\sim 1/3$ solar) should be taken as
representative of the low-$z$ universe as a whole.  There is now
compelling evidence that the bulk of stars not only in cluster
ellipticals but also in field ellipticals and bulges formed at high 
redshifts ($z\gsim 3$). Since such stars account for at least $\sim
30\%$ of the baryons now locked into stars, it is argued that at least
$30\%$ of stars and metals formed before $z\simeq 3$, and
correspondingly the
metallicity of the universe at $z=3$ is predicted to be $\sim 1/10$ solar.
This requires the cosmic star formation rate to run at least flat from 
$z\sim 1$ to $\sim 5$, which appears to agree with the most recent
derect
determinations of the star formation rate in Lyman-break galaxies.
\end{bloisabstract}

\section{Introduction}

The observation of forming galaxies at high redshift is certainly the most
direct way to look at the {\it Birth of Galaxies}, and a great
observational effort is currently being made in this direction. Yet,
high redshift galaxies are very faint, and at least for now only few
of their global properties can be measured. Nearby and moderate
redshift galaxies can instead be studied in far greater detail, and
their {\it fossil evidence} can provide a view of galaxy
formation and evolution that is fully complementary to that given
by high redshift observations. By fossil evidence I mean those
observables that do not refer to ongoing, {\it active} star formation,
and which are instead the result of the integrated past star formation
history.  This evidence includes the global metallicity of the
universe, and the distribution of stellar ages in $z\simeq 0$ galaxies
as well as in high redshift, but passively evolving galaxies.

The paper is organized as follows. Section 2 presents the current
evidence for the chemical composition of local clusters of galaxies
at low redshift, separately for the intracluster medium (ICM) and for
cluster galaxies.
In Section 3 the current evidence  is presented for
the bulk of stars in galactic spheroids (i.e. ellipticals and bulges
alike) being very old, formed at high redshift, no matter whether they
reside in rich clusters or in the low density environment we usually
refer to as the {\it field}. In Section 4 these evidences are used to
set constraints on the main epoch of star formation, metal production,
and on the metallicity of the high-$z$ universe.

\section{Clusters as Archives of the Past Star and Metal Production} 
Individual galaxies are not good examples of the {\it closed box}
model of chemical evolution: enriched matter certainly flows out of
them,
and quasi-pristine material may occasionally infall and be accreted by them.
Theoretical simulations predict instead that the baryon fraction of rich
clusters
cannot change appreciably in the course of their evolution [43]. 
We can then expect within a cluster
to find confined in the same place 
all the dark matter, all the baryons, all the galaxies, and all the
metals that have participated in the play. Clusters are then good {\it
archives} of their past star formation and metal production history,
an ideal laboratory for the study of the fossil evidence.

Metals in clusters are partly spread through their ICM, 
partly locked into galaxies and stars. The mass
of metals in the 
ISM of galaxies is instead negligible compared to that in the two
other components. ICM abundances can be obtained from X-ray 
observations, while optical observations combined to population
synthesis models provide estimates for the metallicity of the stellar
component of galaxies.

\subsection{Iron and $\alpha$-Elements in the Intracluster Medium}

Among abundant elements, iron has the most accurately  measured ICM abundance. 
The
so-called iron-K emission complex at $\sim 7$ keV is prominent in the
X-ray spectrum of clusters, and allows fairly precise measurements. 
Fig. 1 shows the
iron abundance of clusters and groups as a function of ICM
temperature, while  
Fig. 2 shows the iron-mass-to-ligh-ratio (\IMLR) of the ICM [30]. This is
defined as the ratio $\mfecm/\lb$ of the total iron mass in
the ICM over the total $B$-band luminosity of the galaxies in the cluster.

\begin{figure}
\vskip-5.5truecm
\centerline{\psfig{file=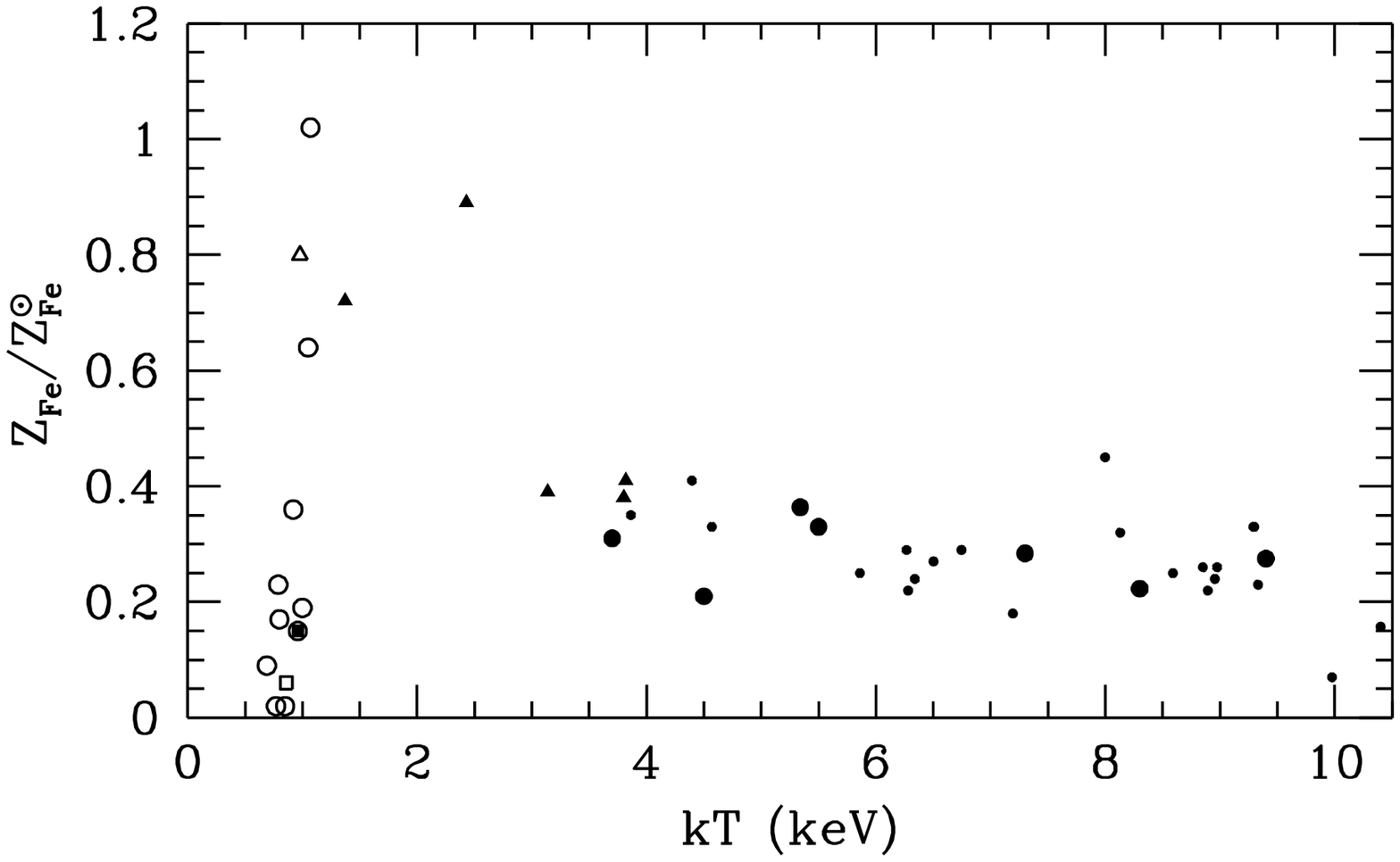,width=15cm,angle=0,height=14cm}}

%\vskip-5truecm
%\plotone{feab.eps}
\vskip-1truecm
\caption{\piedi A compilation of the iron abundance in the ICM as a 
function of ICM temperature
for a sample of clusters and groups [30], including several clusters at
moderately high redshift with $<z>\simeq 0.35$, represented by small
filled circles.}
\end{figure}

The drop of the derived \IMLR \ in poor clusters and groups (i.e. for
 $kT\lsim 2$ keV) can be traced back to a drop in both factors
 entering in its definition, i.e., in the iron abundance (cf. Fig. 1)
{\it and} in
 the ICM mass to light ratio [30].  We don't know whether this drop is a
 real effect.  Groups may not behave as closed boxes, and may be
 subject to baryon and metal losses due to the feedback effect of star
 formation in individual galaxies driving galactic winds [34] [9].  
In addition, there may be a
 diagnostic problem, since in groups iron is derived
 from iron-L transitions, which involve atomic physics which is
 far more complicated than that of the iron-K complex [2][30]. 
For this reason I will not further
 discuss poor clusters, i.e. those cooler than $kT\lsim 2$ keV.

Fig. 1 and 2 show that both the iron abundance and the
\IMLR \ in rich clusters ($kT\gsim 2$ keV) are independent of
cluster temperature, hence of cluster richness and
optical luminosity. For these clusters one has 
$\zfecm =0.3\pm 0.1$ solar, and $\mfecm/\lb = (0.02\pm 0.01)$ for
$\ho=50$.
The most straightforward interpretation is that clusters did not lose
iron (hence baryons), nor selectively acquired pristine baryonic material, and
that the conversion of baryonic gas into stars and galaxies has
proceeded with the same efficiency and stellar IMF in all clusters [30].
Otherwise, we should observe cluster to cluster variations of the iron
abundance and of the \IMLR. 
The theoretical prediction of the constancy of the baryon fraction in clusters
[43] is nicely supported by these evidences. 

\begin{figure}
\vskip-5truecm
\centerline{\psfig{file=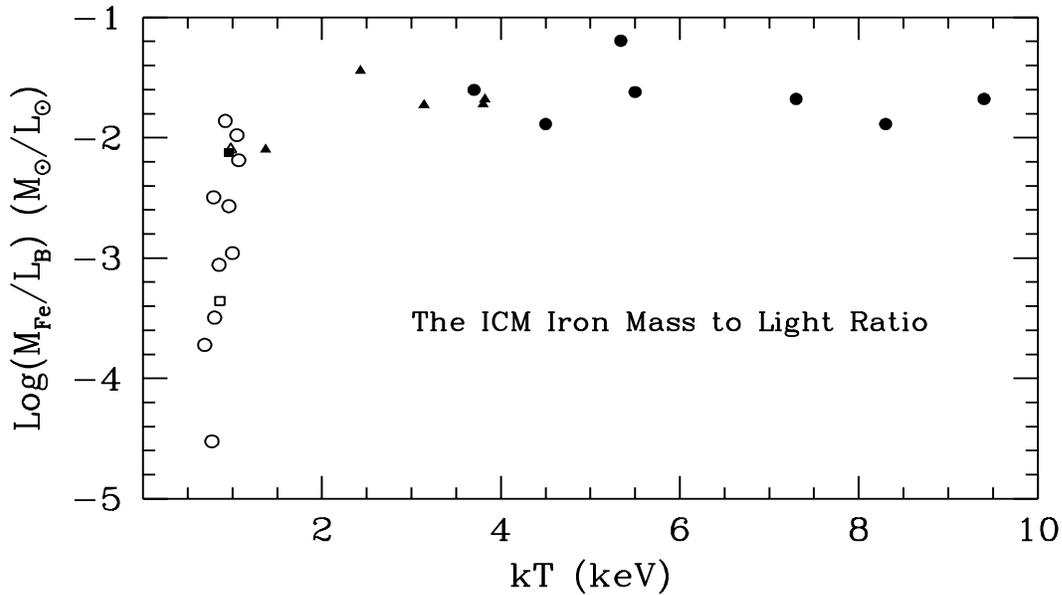,width=15cm,angle=0,height=14cm}}
%\plotone{feml.eps}
\vskip-1truecm
\caption{\piedi The iron mass to light ratio  of  the  ICM of clusters
and groups (for $\ho=50$) as a function of the ICM temperature [30].}
\end{figure}

X-ray observations have also allowed to measure the abundance of
other elements in the ICM, especially that of the $\alpha$-elements such as
O, Ne, Mg, and Si, with the {\it ASCA} X-Ray telescope having
superseded all
previous attempt in this respect. A fairly high $\alpha$-element
enhancement, with
$<\![\alpha$/Fe]$>\simeq +0.4$, was initially reported [23]. Later  this
estimate has been revised down to
$<\![\alpha$/Fe]$>\simeq +0.2$  [24], 
using a global average among observed clusters, and averaging 
O, Ne, Mg, and Si for the $\alpha$-elements [30].
In this way a more robust estimate of the global [$\alpha$/Fe] ratio
is obtained, given the errors affecting the abundances
of individual elements in individual galaxies. This may still suggest a modest
$\alpha$-element enhancement, with the ICM enrichment being dominated
by SNII products [23].

However, it has been pointed out that this 
small apparent $\alpha$-element enhancement in the ICM comes from 
having assumed the reference solar iron abundance from the
``photospheric'' 
model atmosphere analysis [15].
The ``meteoritic'' iron abundance is instead 
now generally adopted, and since this is $\sim 0.16$
dex  lower than the photospheric value, one can conclude that
there is virtually no $\alpha$-element enhancement at all in
the ICM (formally $<\![\alpha$/Fe]$>\simeq +0.04\;\pm\sim 0.2$, [15], [30]).
Clusters of galaxies
are therefore nearly {\it solar} as far as the
elemental ratios are concerned, which argues for stellar 
nucleosynthesis having proceeded in quite the same way in the solar 
neighborhood as well as at the galaxy cluster scale. 
This implies a similar ratio of the number of Type Ia to Type II SNs, 
as well as
a similar IMF [30]. This result speaks for the universality of the
star formation process (IMF, binary fraction, etc.), and
may help limiting the number of free
parameters to play with.

\subsection{The Iron Content of Galaxies and the Iron Share}

The metal abundance of the stellar component of cluster galaxies can
only be inferred from integrated spectra coupled to synthetic stellar 
populations. Much of the stellar mass in clusters is confined to
passively evolving spheroids (ellipticals and bulges) for which the
iron abundance may range from $\sim 1/3$ solar to a few times solar.
Hence, the average iron abundance cannot be much different from solar,
even when taking into account the presence of radial gradients
[2], $\alpha$-element enhancements [8], and the luminosity bias [14].

The  global iron
abundance of a whole cluster is  therefore given by:
$$Z_{\rm CL}^{\rm Fe}={\zfecm\micm + \zfes M_* \over \micm + M_*}=
      {5.5\zfecm h^{-5/2} + \zfes h^{-1}\over 5.5h^{-5/2} +
      h^{-1}},\eqno(1)$$
where $\zfes$ is the average abundance of stars in galaxies and $ M_*$
is the mass in stars. For the second equality 
I have assumed as prototypical the Coma cluster values
adopted by White et al. [43]: $\micm\simeq 5.5\times
10^{13}h^{-5/2}\msun$ and $M_*\simeq 10^{13}h^{-1}\msun$.
With $\zfecm=0.3$ solar and $\zfes=1$ solar,
equation (1) gives a global cluster abundance of 0.34,
0.37, and 0.41 times solar, respectively for $h=0.5$, 0.75, and 1. 
Under the same assumptions, the ratio of the iron mass in the ICM to
      the iron mass locked into stars is:
$${\zfecm\micm\over\zfes M_*}\simeq 1.65 h^{-3/2},\eqno(2)$$ or 4.6,
2.5, and 1.65, respectively for $h=0.5$, 0.75, and 1. Note that with
the adopted values for the quantities in equation (2) most of the
cluster iron resides
in the ICM, rather than being locked into stars, especially for low
values of $\ho$. These estimates could be somewhat decreased if
clusters contain a sizable population of stars not bound to censed
individual galaxies, if the average iron abundance in stars is
supersolar (luminosity-weighted determinations underestimate true
abundances [14], or if the galaxy $M_*/L$ ratio is higher
than adopted here, i.e., $<\!M_*/\lb\!>=6.4h$ [43].
However, the bottom line is that there is at least as much metal mass inside
cluster galaxies, as there is out of them in the ICM. This must be
taken as
a strong constraint by models of the chemical evolution of galaxies:
clearly galaxies do not evolve as a closed box, and outflows must play a
leading role.

With the adopted masses and iron abundances for the two baryonic
components  one can also evaluate the total
cluster \IMLR:
$${\mfecm +\mfes\over\lb}\simeq 1.3\times 10^{-2}(1.65\, h^{-1/2}+h)
\; (\msun/\lsun),\eqno(3)$$
or \IMLR=0.037 or 0.034 $\msun/\lsun$, respectively for $h=0.5$ and
1. The total
\IMLR \ is therefore fairly insensitive to the adopted distance scale.
Simple calculations 
show that to reproduce this value one needs either
a fairly flat IMF ($x\simeq 0.9$) if all iron is attributed to SNII's,
or a major contribution from SNIa's, if one adopts a Salpeter IMF
($x=1.35$) [34]. The former option
dictates a substantial $\alpha$-element enhancement, similar to the
values observed in the Galactic halo ([$\alpha$/Fe]$\simeq +0.5$).
The latter option instead predicts near solar proportions for the
cluster as a whole, and requires the SNIa rate to have been much
higher in the past [34]. The evidence presented in Section 2.1  favors 
the second option. 

From the near solar proportions of cluster
abundances one obtains the total metal mass to light ratio of a typical
cluster as $M_{\rm Z}/\lb\simeq 10\times M_{\rm Fe}/\lb\simeq  0.3\pm 0.1\;
(\msun/\lsun)$.
It is worth noting that this is an interesting, fully empirical
estimate of the metal
yield of stellar populations. Following
Tinsley [41], the metal yield is usually defined per unit mass of
stars, a quantity which theoretical counterpart depends on the poorly known
low mass end of the
IMF. The estimate above gives instead the yield per unit luminosity of
present day cluster galaxies, a quantity that depends on the IMF only
for $M\gsim\msun$.  Theoretical mass-related yields have been recently
estimated by Thomas, Greggio, \& Bender
[40] based on massive star models and supernova explosion [39], [44]. 
These yields can be purged
from their mass dependence, and transformed into luminosity-related
yields. For this purpose I assume an age of 15 Gyr for the bulk of
stars in clusters (cf. Section 4), and use the proper luminosity-IMF
normalization [29] [31]: i.e. $\psi(M)=AM^{-(1+x)}$ for the IMF with
$A\simeq 3.0\lb$. Thus, theoretical yields turn out to be $M_{\rm
Z}/\lb=0.08$, 0.24, and $0.33\; \msun/\lsun$, respectively for
$x=1.7$, 1.35, and 1.00, which compares to $M_{\rm Z}/\lb\simeq 0.3\pm
0.1 \msun/\lsun$ for the empirical cluster value. One can conclude
that current stellar yields do not require a very flat IMF to account
for the cluster metals.

\subsection{Clusters as Fair Samples of the Local Universe}

To what extent the cluster global metallicity, and
the ICM to galaxies iron share are representative of the low-$z$
universe as a whole? For example, Madau et al. [21] adopt $\ho=50$, a stellar
mass density parameter $\Omega_*=0.0036$, and a baryon mass density
parameter $\Omega_{\rm b}=0.05$. With these values the fraction of
baryons that have been locked into stars is $\sim 7\%$. 
This compares to  $\sim
1/(1+5.5h^{-3/2})$ in clusters, or $\sim 6\%$ and $\sim 10\%$,
respectively for $h=0.5$ and 0.75.
Therefore, it appears that the efficiency of baryon conversion into
galaxies and stars adopted  by Madau et
al. [21] is nearly the same as that observed in clusters, which
supports the notion of clusters being representative of the low-$z$
universe $(\Omega_*/\Omega_{\rm b})_\circ$. 

For lack of direct evidence on the metallicity of the local
 intergalactic medium (IGM),  Madau et
al. [21] assumed all the metals in the low redshift universe to be  
 locked into stars (galaxies),  with
a negligible metal content for the IGM, which however
comprises the vast majority of all the baryons. With these
assumptions, and adopting the average metallicity of all stars to be
 solar,  the metallicity of the present day universe is $\sim
1\times 0.0036/0.05= 0.07$ solar, or $\sim 5$ times lower than the measured
value in clusters of galaxies. 

Is this difference real, or does it just reflect the assumption of a
zero metallicity IGM? There is no obvious reason why the metal yield
of stellar populations should be $\sim 5$ times lower in field
galaxies compared to cluster galaxies.
On the contrary, assuming the
difference to be real would force us to accept
a drastic difference between the behavior of
 galaxies and stellar populations in  clusters and in the field, which
 would require quite contrived explanations [30]. Much simpler, hence
more attractive, appears to be the option according to which no major 
difference in metal enrichment exists between
 field and clusters, and  the global
metallicity of the present day universe is nearly the same as
that of the only place where we can thoroughly measure it, i.e. in galaxy
clusters where it is $\sim 1/3$ times solar.
If so, there should be a comparable share of metals in the IGM,
as there is in the cluster ICM, i.e., just like most of the baryons,
most of the metals should reside
in the IGM rather than within field galaxies [30].
\vfil\newpage

\section{The Age of Spheroids and the Main Epoch of Metal Production}

The cluster abundances as illustrated in the previous section don't say
much about the cosmic epoch when the bulk of the cluster metals were
produced and dispersed through the ICM. The only constraint comes from the
iron abundance in moderate redshift clusters ($z\simeq 0.5$) being
virtually identical to that of local clusters (see Fig. 1), 
hence the bulk of iron had to be manufacted at $z\gsim 0.5$.
A much more stringent constraint comes from current age estimates of
the dominant stellar populations in cluster ellipticals, the likely
producers of the bulk of the metals, and from other
fossil evidences which are schematically reported in this section.

A first tight constraint on the formation epoch of stars in cluster
ellipticals came from Bower, Lucey, \& Ellis [6] who noted that the very tight 
color-$\sigma$ relation followed by galaxies in the Virgo and Coma
clusters demostrates that at least
{\it cluster} ellipticals are made of very old stars, with the bulk
of them having formed at $z\gsim 2$. 
This result had the merit to cut short inconclusive discussions on the
age of ellipticals based on matching synthetic spectra to individual
galaxies, and showed instead that the homogeneity of elliptical
populations sets tight, almost model independent age constraints.
Along these lines, evidence supporting an early formation of ellipticals
 has greatly expanded over the last few years. This
came from the tightness of the fundamental plane
relation for ellipticals in local clusters [33],
from the tightness of the color-magnitude relation for ellipticals in
clusters up to $z\sim 1$ (e.g., [1], [37]), and from the
modest shift with increasing redshift in the zero-point of the fundamental
plane, Mg$_2-\sigma$, and color-magnitude relations of cluster
ellipticals (e.g., [4], [10], [12], [18], [26], [37], [42]).
All these studies
agree in concluding that most stars in cluster ellipticals formed at $z\gsim
3$, though the precise value depends on the adopted cosmological
parameters (e.g. in a $\Lambda$-dominated universe this constraint
would become $z\gsim 2$). 

\begin{figure}
\vskip-1truecm
\centerline{\psfig{file=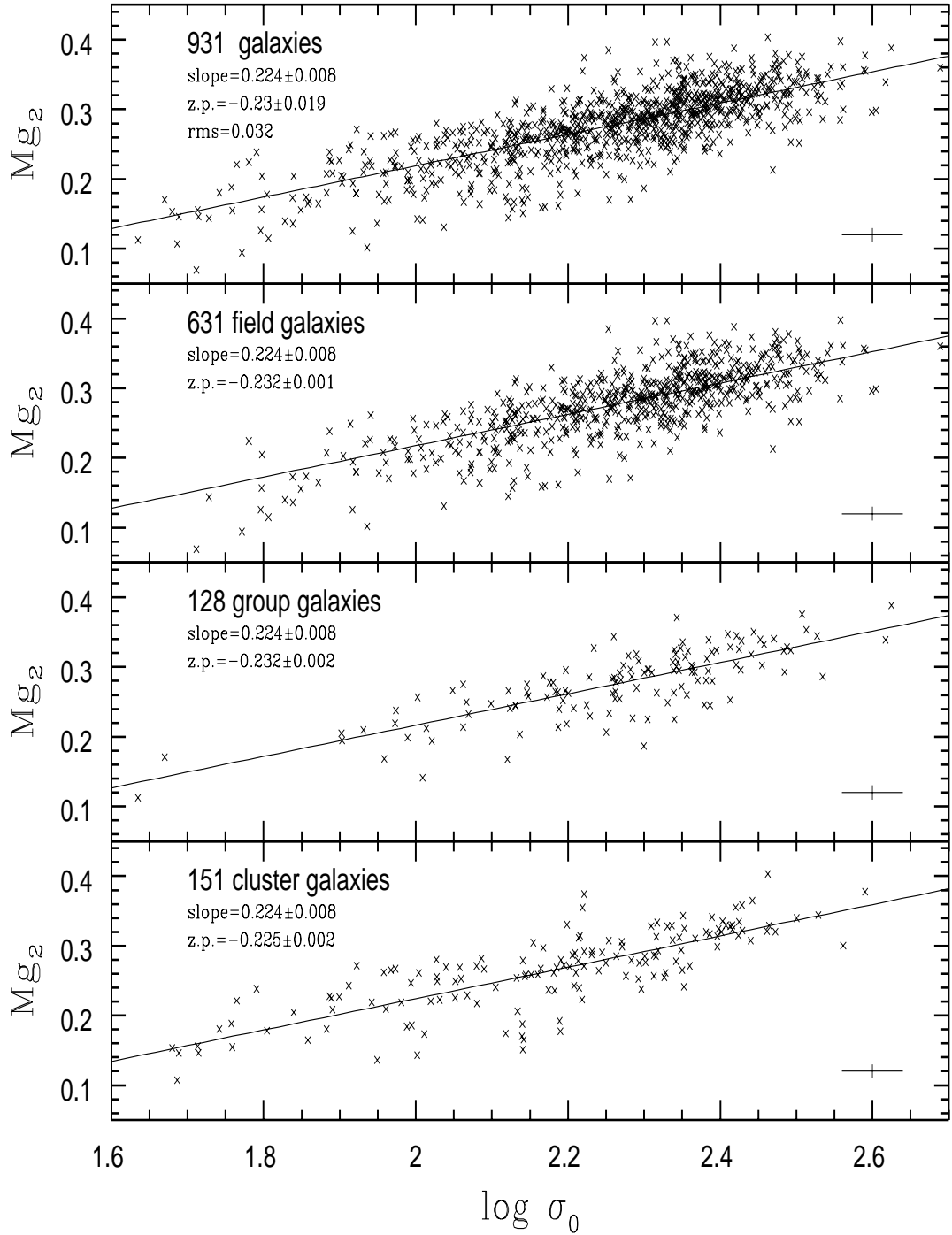,width=15cm,angle=0,height=20cm}}

%\vskip-5truecm
%\vskip-1truecm
\caption{The Mg$_2-\sigma$ relation for a sample of early-type
galaxies (upper panel), as well as for the field, group and cluster
subsamples (lower panels). The corresponding number of objects,
the slope, and the zero-point (z.p.) are shown in the upper
left corner of each panel. he least squares fits to the Mg$_2-\sigma$ relation
are also shown. Note that for the three subsamples the slope as
derived for the total sample was retained, and only the zero-point was
determined [5].}
\end{figure}

This cogent  result has been established for cluster ellipticals, but
additional fossil evidence argues for its validity for field
ellipticals as well as for most bulges of spirals, i.e. in general
for the vast majority of galactic {\it spheroids}.  Indeed, as shown
in Fig. 3, field early-type galaxies (ellipticals and S0's) are found
to follow virtually the same Mg$_2-\sigma$ relation of their cluster
counterparts, with the age difference being less than $\sim 1$ Gyr
[5], while most bulges follow the same
Mg$_2-\sigma$ and color-magnitude relations of ellipticals [17]. 
Even more directly, the bulge of our own
Galaxy is found to be dominated by stars which age is
indistinguishable from that of the Galactic halo, or $\gsim 12$ Gyr
[25], and similar conclusions have been reached for 
the bulge of M31 [31], [16]. 
 It is also worth noting that the close
resemblance of cluster and field early-type galaxies illustrated in
Fig. 3 indicates that metal enrichment is not detectably
different in cluster and in field galaxies.

 With spheroids containing at least 30\% of all
stars in the local universe [27], [35], or even more [13], one can
conclude that at least
 30\% of all stars and metals have formed at $z\gsim 3$ (Renzini [32];
see also Dressler \& Gunn [11]).
This is several times more than suggested by a conservative
 interpretation of the early attempt at tracing the cosmic history of
 star formation, either empirically [21] or from 
theoretical simulations, e.g., [3]. 
Yet, it is is more in line with
recent direct estimates from the spectroscopy of Lyman-break galaxies
[38], where the cosmic SFR runs flat for $z\gsim 1$,
as in one of the options offered by the models of Madau et al. [22].

At first sight these inferences from the local, fossil evidence appear
to conflict with some model predictions, and with some of the current
 interpretations of the direct observations at $z\gsim
3$.  For example, the standard CDM model of the Durham group predicts
 that only $\lsim 5\%$ of stars have
formed by $z=3$ [3], [7]. The supernova-feedback free parameter of this
specific CDM model was tuned to reproduce a $z=0$ galaxy luminosity
function that runs flat at the faint end. Tuning instead to the Zucca
et al. [45] luminosity function, which is steep at the faint end,
allows to make much more stars at early times (Frenk,
this conference), in somewhat better agreement
with the constraint sets by the old age of galactic spheroids.

With $\sim 30\%$ of all stars having formed at $z\gsim 3$, and the
metallicity of the $z=0$ universe being $\sim 1/3$ solar, it is
straightforward
to conclude that the global metallicity of the $z=3$ universe had to
be $\sim 1/3\times 1/3\sim 1/10$ solar,
or more [32]. Damped Ly$_\alpha$
systems (DLA) may offer an opportunity to check this prediction, though they
may provide a vision of the early 
universe that is biased in favor of cold, metal-poor gas that has been
only marginally affected by star formation and metal pollution.
Metal rich objects such as giant starbursts that would be dust obscured, 
the metal rich
passively evolving spheroids, and the hot ICM/IGM obviously do not
enlist among DLAs.
Yet, in spite of these limitations the average metallicity of the DLAs
at $z=3$ appears to be
$\sim 1/20$ solar (Pettini et al. [28], see their Fig. 4), just a factor of
2  below the expected value from the {\it fossil evidence}. 
This is still much higher than the extreme lower limit 
$Z\simeq 10^{-3}Z_\odot$ at $z=3$ as
inferred from Ly$_\alpha$ forest observations [36]. Ly$_\alpha$ forest material
is believed to contain a major fraction of cosmic baryons at high $z$, hence
(perhaps) of metals. There is therefore a potential conflict with
the estimated global metallicity at $z\simeq 3$, and the notion of 
Ly$_\alpha$ forest metallicity being representative of the the
universe metallicity at this redshift.
Scaling
down from the cluster yield, such low metallicity was achieved when
only $\sim 0.3\%$ of stars had formed, which may be largely
insufficient to
ionize the universe and keep it ionized up to this redshift [20]. 
This suggests that Ly$_\alpha$ forest may not trace the 
mass-averaged metallicity
of high redshift universe, and that the universe was very
inhomogeneous at that epoch. The bulk of metals would be partly locked
into stars in the young spheroidals, partly would reside 
in a yet undetected hot
IGM, a phase hotter than the Ly$_\alpha$ forest phase.
\par\bigskip
 \noindent
I would like to thank Len Cowie and Sandra Savaglio for useful
 discussions on the chemestry of DLAs and the Ly$_\alpha$ forest.


\begin{bloisbib}
\bibitem{} Aragon-Salamanca, A., Ellis, R.S., Couch, W.J. \&
     Carter, D. 1993, MNRAS, 262, 764 
\bibitem{} Arimoto, N., Matsushita, K., Ishimaru, Y., Ohashi, T., \&
    Renzini, 1997, ApJ, 477, 128 
\bibitem{} Baugh, C.M., Cole, S.,  Frenk, C.S., \& Lacey, C.G. 1997,
    Astro-ph 9703111
\bibitem{} Bender, R., Saglia, R.P., Ziegler, B., Belloni, P., Greggio, L.,
     Hopp, U. \& Bruzual, G.A. 1997, ApJ, 493, 529
\bibitem{} Bernardi, M., Renzini, A., da Costa, L.N., Wegner, G.,
           Alonso, M.V., Pellegreini, P.S., Rit\'e, C., \& Willmer,
           C.N.A. 1998, ApJL, in press (astro-ph/9810066)
\bibitem{} Bower, R.G., Lucey, J.R. \& Ellis, R.S. 1992, MNRAS, 254, 613
\bibitem{} Cole, S., Aragon-Salamanca, A., Frenk, C.S., Navarro, J.S., \&
    Zepf, S.E. 1994, MNRAS, 271, 781
\bibitem{} Davies, R.L., Sadler, E.M., \& Peletier, R.F. 1993, MNRAS,
    262, 650
\bibitem{} Davis, D.S., Mulchaey, J.S., \& Mushotzky, R.F. 1998, 
           astro-ph/9808085
\bibitem{} Dickinson, M. 1995, in Fresh Views of Elliptical Galaxies,
         ed. A. Buzzoni, A. Renzini, \& A. Serrano, ASP Conf. Ser. 86,
         283
\bibitem{} Dressler, A., \& Gunn, J. E. 1990, ASP. Conf. Ser. 10, 204
\bibitem{} Ellis, R.S., Smail, I., Dressler, A., Couch, W.J., Oemler,
     A. Jr., Butcher, H., \& Sharples, R.M. 1997, ApJ, 483, 582
\bibitem{} Fukujita, M., Hogan, C.J., \& Peebles, P.J.E. 1998, 
          astro-ph/9712020v2
\bibitem{} Greggio, L. 1997, MNRAS, 285, 151
\bibitem{} Ishimaru, Y., \& Arimoto, N. 1997, PASJ, 49, 1
\bibitem{} Jablonka, P., Bridges, T.J., Sarajedini, A., Meylan, G.,
         Maeder, A., \& Meynet, G. 1998
\bibitem{} Jablonka, P., Martin, P., \& Arimoto, N. 1996, AJ, 112, 1415
\bibitem{} Kodama, T., Arimoto, N., Barger, A.J., \&
      Aragon-Salamanca, A. 1998, astro-ph/9802245
\bibitem{} Kauffmann, G. 1996, MNRAS, 281, 487
%\reference Kodama, T., \& Arimoto, N. 1997, A\&A, 320, 41
\bibitem{} Madau, P. 1998, astro-ph/9807200
\bibitem{} Madau, P., Ferguson, H.C., Dickinson, M.E., Giavalisco, M.,
     Steidel, C.C., \& Fruchter, A. 1996, MNRAS, 283, 1388
\bibitem{} Madau, P., Pozzetti, L., \& Dickinson, M. 1997, Astro-ph 9708220
\bibitem{} Mushotzky, R.F. 1994, in Clusters of Galaxies, ed. F. Durret,
     A. Mazure, \& J. Tran Thanh Van (Gyf-sur-Yvette: Editions
     Fronti\`eres), p. 167
\bibitem{} Mushotzky, R.F., et al. 1996, ApJ, 466, 686
\bibitem{} Ortolani, S., Renzini, A., Gilmozzi, R., Marconi, G., Barbuy, B.,
     Bica, E., \& Rich, R.M., 1995, Nat, 377, 701
\bibitem{} Pahre, M.A., Djorgovski, S.G., \& de Carvalho, R.R. 1997, in
     Galaxy Scaling Relations: Origins, Evolution and Applications,
     ed. L. da Costa \& A. Renzini (Berlin: Springer), p. 197
%\bibitem{}Peebles, P.J.E. 1997, ApJ, 483, L1
\bibitem{} Persic, M., \& Salucci, P. 1992, MNRAS, 258, 14p
\bibitem{} Pettini, M., Smith, L.J., King, D.L., \& Hunstead,
     R.W. 1997, ApJ, 486, 665
\bibitem{} Renzini, A. 1994, Galaxy Formation, ed. J. Silk \&
     N. Vittorio (Amsterdam: North Holland), p. 303
\bibitem{} Renzini, A. 1997, ApJ, 488, 35 
\bibitem{} Renzini, A. 1998a, AJ, 115, 2459
\bibitem{} Renzini, A. 1998b, astro-ph/9801209, in The Young Universe,
     ed. S. D'Odorico, A. Fontana, E. Giallongo, ASP Conf. Ser. 146, 298
\bibitem{} Renzini, A., \& Ciotti, L. 1993, ApJ, 416, L49
\bibitem{} Renzini, A., Ciotti, L., D'Ercole, A., \& Pellegrini, 
     S. 1993, ApJ, 419, 52
\bibitem{} Schechter, P.L., \& Dressler, A. 1987, AJ, 94, 563
\bibitem{} Songaila, A. 1997, ApJ, 490, L1
\bibitem{} Stanford, S.A., Eisenhardt, P.R., \& Dickinson, M. 1998,
           ApJ, 492, 461
\bibitem{} Steidel, C.C., Adelberger, K.L., Giavalisco, M., Dickinson,
           M., \& Pettini, M. 1998, preprint
\bibitem{}  Thielemann, F.-K., Nomoto, K., \& Hashimoto, M. 1996, ApJ,
     460, 408
\bibitem{}  Thomas, D., Greggio, L., \& Bender, R. 1997, Astro-ph 9710004
\bibitem{}  Tinsley, B.M. 1980, Fund. Cosmic Phys. 5, 287
\bibitem{} van Dokkum, P. G., Franx, M., Kelson, D. D.,
     \& Illingworth, G. D. 1998, ApJ, 504, L17
\bibitem{} White, S.D.M., Navarro, J.F., Evrard, A.E., \& Frenk,
     C.S. 1993,   Nat, 366, 429
\bibitem{} Woosley, S.E., \& Weaver, T.A. 1995, ApJS, 101, 181
\bibitem{} Zucca, E., et al. 1997, A\&A, 326, 477
\end{bloisbib}
\end{document}